# Beyond Productivity: Measuring Developers' Cognitive Load During GenAI-Supported Software Development

Charlotte Brandebusemeyer
*Digital Health – Connected Healthcare*
*Hasso Plattner Institute, University of Potsdam*
Potsdam, Germany
char.brandebusemeyer@hpi.de

Daniela Gasser
*Digital Health – Connected Healthcare*
*Hasso Plattner Institute, University of Potsdam*
Potsdam, Germany
daniela.gasser@guest.hpi.de

Tobias Schimmer
*SAP Labs*
*SAP*
Newport Beach, USA
tobias.schimmer@sap.com

Bert Arnrich
*Digital Health – Connected Healthcare*
*Hasso Plattner Institute, University of Potsdam*
Potsdam, Germany
bert.arnrich@hpi.de

***Abstract*— Generative AI (GenAI) is changing software development workflows and how developers work. Industry evaluations of GenAI adoption often monitor productivity gains, usage, and output quality, but limited attention is paid to the interaction experience and cognitive load of the actual adopters and drivers of GenAI technology – the software developers. Understanding whether GenAI changes or shifts developers' cognitive demands during everyday development is important for a developer-centered evaluation of GenAI-supported software development. It can inform organizations in designing and evaluating effective AI-supported workflows. In this work, we study how GenAI use and task context relate to professional developers' perceived cognitive load and whether wearable-derived physiological characteristics provide additional information beyond this context. In a four-day industrial field study at two SAP sites, 21 developers documented their tasks, task duration, GenAI use, and perceived cognitive load while wearing an EmbracePlus wristband. The results show that perceived cognitive load is associated with both GenAI use and task context, while physiological measures provide only limited additional information. These findings suggest that developers' perceived cognitive load during GenAI-supported software development should be evaluated in relation to the concrete work context, with wearable physiological data used as complementary rather than standalone information.**

## I. INTRODUCTION

Generative AI (GenAI) tools are increasingly integrated into professional software development workflows. Organizations commonly monitor GenAI adoption through active user rates, usage frequency, prompt or code completion counts, and suggestion acceptance counts, while productivity effects are often assessed through task completion time, number of completed work items, pull-request cycle times and software delivery metrics. However, these measures offer limited insight into the cognitive demands developers face when using GenAI in everyday software development. The evaluation of AI-generated suggestions and associated code comprehension places substantial demands on developers' cognitive resources, with potential implications for task performance [17, 32], sustained effective work and the broader developer experience. Understanding the factors associated with perceived cognitive load and how to measure it during AI-supported software development in everyday real-life work settings is relevant for evaluating AI tools, designing effective workflows, and supporting long-term developer well-being.

Previous analyses of this broader study examined developers' subjective and behavioral experiences with GenAI [8, 10], suggesting that cognitive load may depend on how developers interact with GenAI and evaluate its output. However, these analyses did not jointly estimate the associations of GenAI use, task category, and task duration with perceived cognitive load. This combined analysis is important because cognitive demands during AI-supported work may reflect both the task itself as well as the additional effort required to interact with and supervise the AI system.

Measuring cognitive load during everyday software development is challenging. Questionnaires can provide task- and context-specific information about developers' perceived cognitive load. However, self-reported ratings are subject to individual differences in how scales are used and interpreted, as well as in individual self-assessment and introspection. Such subjective ratings lack an objective baseline and can interrupt workflows. Wearables offer a less intrusive way to collect continuous, objective physiological data, such as cardiovascular activity, electrodermal activity, and skin temperature, which may provide complementary indicators associated with cognitive load. Controlled studies have shown that physiological responses measured with wearables can distinguish experimentally induced levels of cognitive load [2, 4, 19, 35]. However, physiological signals in real-world work settings are affected by multiple non-cognitive influences, such as movement, device fit, environmental conditions, the type of activity performed, and affective states. Thus, it remains unclear how informative individual wearable-derived physiological metrics and multimodal physiological characteristics are for understanding developers' perceived cognitive load in practice.

Our previous work has shown that wearable-derived physiological data can be collected during developers' everyday work and may relate to perceived cognitive load [7]. The present study extends this work by examining whether individual wearable-derived physiological metrics or multimodal physiological characteristics provide detectable information about perceived cognitive load once GenAI use, task category, task duration, and participant-specific differences are accounted for.

We investigated these aspects in a real-world industrial setting with professional software developers at two SAP sites. During the study, developers documented their everyday work tasks, the start and end times of each task, whether GenAI was used, and their perceived cognitive load during each task. In parallel, physiological activity was recorded using an EmbracePlus wristband. This setup allowed us to examine perceived cognitive load during developers' natural work activities rather than only during controlled laboratory tasks, while linking task-level subjective ratings, task context, GenAI use, and wearable-derived physiological data.

We address the following research questions in this work:

**RQ1:** Is GenAI use associated with developers' perceived cognitive load after accounting for the working task category and task duration, and how are these task characteristics themselves associated with perceived cognitive load during everyday software development?

**RQ2:** Do wearable-derived physiological characteristics, individual or multimodally, provide additional information about variations in perceived cognitive load beyond GenAI use, task category, and task duration during everyday software development tasks?

Our contribution is a context-aware and interpretable analysis of how GenAI use, task context, and individual and combined wearable-derived physiological parameters relate to perceived cognitive load during real-world software development. We contribute a real-world industrial analysis of perceived cognitive load during GenAI-supported software development and derive practical implications for using wearable data in real-world software engineering settings.

## II. Background and Related Work

### A. Cognitive Load in GenAI-Supported Software Development

Software developers are one of the greatest adopters and drivers of GenAI technology. It is therefore important to take a developer-centered perspective on development productivity that can complement traditional productivity measures by considering how developers experience AI-supported work. The developer experience framework [24] identifies three core dimensions for a developer-centered perspective on productivity and work satisfaction: cognitive load (the amount of mental processing required to perform a task), feedback loop (the effectiveness and efficiency of interactions between developers and also between developers and tools) and flow (a positive mental state of optimal productivity). In this work, we focus on the cognitive load aspect of developer experience in the context of GenAI-supported software development.

In AI-supported development, cognitive load can arise not only from the underlying software engineering task but also from interacting with the AI assistant. Developers need to choose an appropriate interaction style, formulate prompts, comprehend generated code, evaluate its correctness, and integrate it into the existing codebase. It is therefore useful to examine whether perceived cognitive load is associated with the underlying task context, with GenAI use, or with both. If GenAI use remains associated with perceived cognitive load beyond the task context, this would motivate the design of interaction mechanisms that reduce avoidable AI-related cognitive demands and help developers direct more cognitive resources to the working task content itself.

Empirical studies on the cognitive demands of AI-assisted programming are as yet limited and mostly based on controlled setups. Collectively, they suggest that developers spend substantial cognitive effort in understanding and validating AI-generated code rather than generating it. Difficulties can arise in understanding, editing and debugging AI-generated solutions [32]. Likewise, awareness that code had been generated by AI increased developers' effort to inspect the code and locate potential errors, and increased their perceived cognitive workload [30]. While these studies have provided valuable insights into AI's impact on cognitive demands, the predefined tasks and controlled study environments limit the transferability to everyday development work.

Industrial research on AI coding assistants has primarily focused on productivity gains, AI tool adoption outcomes, and developers' perceptions of AI-supported work. Studies with professional developers have reported higher task completion rates with GitHub Copilot [14], reduced task completion time [26], productivity benefits that vary across users [36], and changes in developers' perceptions of AI usefulness, enjoyment, trust and everyday work practices [11]. However, these studies do not explicitly measure cognitive load during naturally occurring work activities.

Only a small number of studies have adopted a more explicit developer-centered perspective on professional developers' cognitive load during real-world AI-supported software development. Vella and Blincoe conducted a longitudinal, questionnaire-based study and found that developers generally perceived AI coding assistants as beneficial for cognitive load and perceived productivity [33]. Previous analyses of the controlled setup part of the broader field study dataset used in this work showed that developers' perceived workload depends not only on general GenAI use but also on the type and intensity of AI interaction for a specific working task [8]. Analyses of the uncontrolled working tasks of this dataset indicated that perceived cognitive load is linked more to the AI interaction, while perceived productivity depends more on the AI output quality [10]. Together, these findings motivate a detailed analysis of how GenAI use and task context are associated with developers' perceived cognitive load during everyday AI-supported software development in an industrial setting.

### B. Wearable-Based Measurement of Cognitive Load

Cognitive load is most commonly assessed using subjective questionnaires, with the NASA Task Load Index (NASA-TLX) being one of the most widely used [20]. Questionnaires provide direct insights into developers' experiences, but they rely on self-assessment and may interrupt ongoing work activities. Prior and current research have therefore investigated whether physiological activity can provide objective indicators for cognitive load. This data could capture objective, continuous, real-time indicators of a person’s experience. Controlled experiments using standardized cognitive tasks, such as the n-back and Stroop paradigms, have shown that cardiovascular measures, such as heart rate and heart rate variability, and electrodermal activity, can distinguish experimentally induced cognitive workload conditions [2, 27]. However, physiological responses are not specific to cognitive load and can also reflect physical movement, affective state, environmental conditions, and individual physiological variation. Collecting physiological signals from minimally invasive devices, such as wearable wristbands, in real-world settings with all the surrounding confounding factors and deducing cognitive load from the signals is an ongoing challenge, which we want to further investigate in this work.

Within software engineering, neurophysiological studies have mostly examined code comprehension, task difficulty, and related developer states under controlled conditions [9, 35]. Work by Fritz et al. showed that multimodal physiological sensing can be used to distinguish developers' cognitive states and task difficulty during programming tasks [18]. For the field study in this work, we deduced that rather than solely relying on individual physiological signals to infer cognitive states in confound-heavy everyday settings, combining physiological recordings with subjective questionnaires and behavioral measures could give holistic insights into cognitive load during software development.

A previous field study with professional software developers reported exploratory associations between perceived cognitive load, working task categories, and wearable-derived physiological activity [7]. Complementary work on that same dataset subsequently evaluated whether multimodal physiological signals and task category information could predict perceived cognitive load, with personalized models showing the strongest performance [31]. Together, these studies establish the feasibility and predictive potential of wearable sensing in real-world software engineering. However, existing research has not yet jointly investigated the associations of GenAI use, task characteristics (e.g., task category and task duration), and wearable-derived physiological measures with perceived cognitive load.

## III. Methods

### A. Participants

The participant cohort [6, 8, 10], comprised 22 SAP employees recruited from two company sites in California, USA. The sample included 12 software developers/engineers, 8 senior software developers/engineers, one senior quality specialist, and one principal software architect. Participants reported between 5 and 10 years of experience in software development and related IT disciplines on average. Eligibility for participation required prior experience with Java, and all participants reported regular use of Java in their work. Python and JavaScript were also commonly used programming languages. Self-assessments indicated a high level of proficiency in Java and proficiency with GenAI tools. GitHub Copilot and ChatGPT were the most frequently used GenAI tools in professional settings, whereas ChatGPT and Gemini were most commonly used outside of work. Most participants reported having used GenAI tools for 6–12 months, with all participants having had at least 1–6 months of prior experience. Due to non-adherence to the study protocol, one participant is excluded from the current analyses. To preserve participant privacy, no demographic information regarding age or gender was collected. Participation was voluntary, and the study design and procedures received approval from the ethics committee of the University of Potsdam.

### B. Study Design and Procedure

In Q2 2025, professional software developers from SAP participated in this four-day study. During the study period, they documented their workday using a workday questionnaire, and their physiological activity was recorded (Fig. 1). In addition to the real-world workplace study period, the developers also participated in controlled study sessions with predefined tasks on the first and last study days. These sessions are outside the scope and focus of this paper. Details on the entire study procedure and findings from the controlled sessions are presented in [1] and [10].

During the uncontrolled study phase, work activities were documented using a paper-based questionnaire. The pen-and-paper format was chosen instead of an online version to provide a continuous visual reminder for participants to fill in the questionnaire without interrupting their workflow. Each working task, its start and end times, associated perceived cognitive load (7-point Likert scale) and productivity (6-point Likert scale), GenAI tools used for the task (yes/no), and the perceived helpfulness of a GenAI tool if used (open-ended question) were documented (Fig. 1). The participants were encouraged to interact with GenAI tools for at least one continuous hour per day. They were free to choose which GenAI tool to use.

In addition to subjective questionnaire data, continuous physiological data were gathered throughout the workday. Participants were given an EmbracePlus wristband [16] (Fig. 2), a smartphone to pair with for data recording, and written detailed instructions on how to record data (Fig. 1). At the start of each workday, participants logged into the “Care Lab App” on their smartphones by scanning a QR code with their participant credentials. The smartphone was then paired with the wristband via Bluetooth. Participants were instructed to wear the wristband on their non-dominant hand and to ensure skin contact with the sensors to ensure reliable measurements. At the end of the workday, the participants synchronized the data between the wristband and the smartphone and logged out of the application on the smartphone. The physiological data could be accessed by the experimenter via the Care Lab Portal provided by the wristband manufacturer, Empatica.

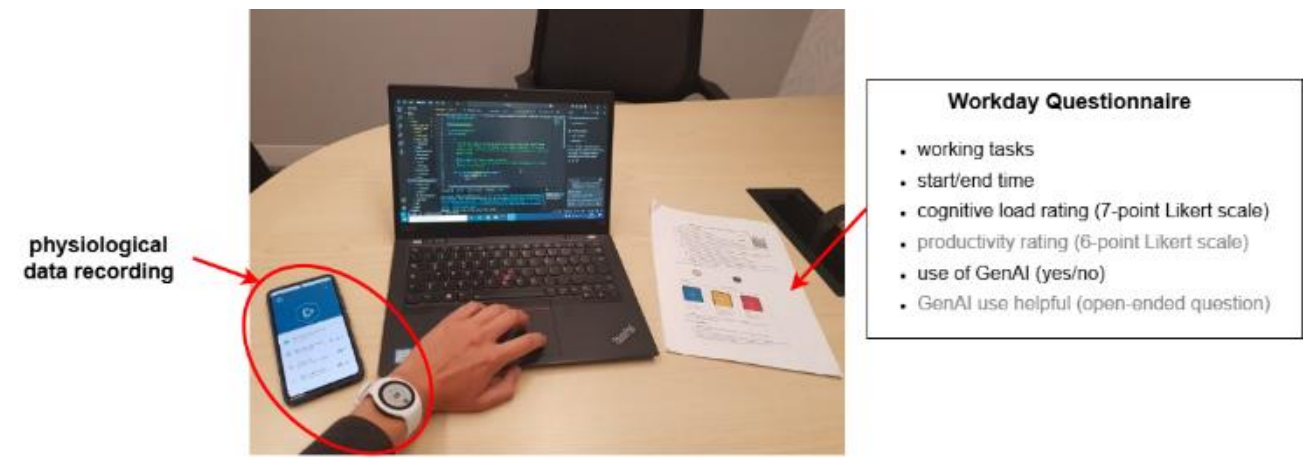

Fig. 1. An exemplary participant’s workplace setup: Whilst working, the paper-based workday questionnaire (with also wristband and application instructions attached) is in visual distance and is filled in. The wristband records physiological data while being paired to the application on the smartphone. The black workday questions are those relevant in this present paper, while the light grey workday questions are only displayed for completion, but are not the focus of the present paper. The figure was taken and adapted from [8].

### C. Working Task Categorization

After the study, the participants' documented working tasks were grouped into three main categories: development-heavy, collaboration-heavy or other activities. The categorizations were conducted by the authors of [10] and are based on those of [22] and [7].

### D. Wearable Device

The EmbracePlus wristband (Fig. 2) was used in this study to record participants' physiological activity. It is equipped with sensors that measure blood volume pulse (BVP) at 64Hz, from which heart rate and heart rate variability can be derived; electrodermal activity (EDA) at 4Hz; skin temperature (TEMP) at 1Hz; and 3-axis wrist acceleration (ACC) at 64Hz [16].

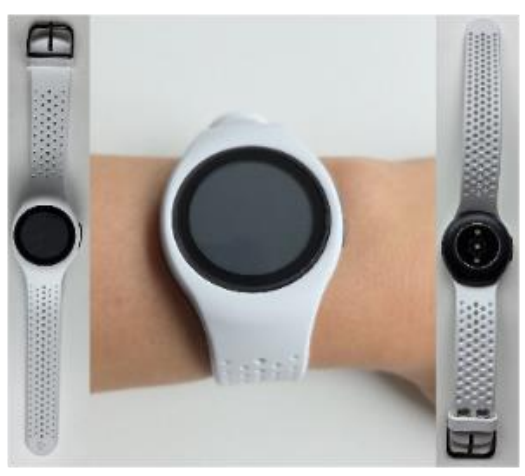

Fig. 2. EmbracePlus wristband used in this study to record participants' physiological activity.

### E. Physiological Signal Processing

Prior to statistical analyses, the physiological signals were pre-processed to reduce measurement artefacts, e.g. produced due to movement, and to derive reliable physiological metrics. Signal processing was performed separately for each participant and working day. Detailed descriptions of pre-processing steps per physiological signal, and the evaluation and selection of the pre-processing pipeline are provided in the supplementary material [5].

#### 1) Acceleration (ACC)

Wrist acceleration data were used solely to detect motion-related artefacts in the physiological recordings and were not included in the statistical analyses. For each participant and working day, a motion threshold was calculated. This threshold was used to identify and remove high-motion segments from the physiological recordings before further processing. The selected thresholding approach was chosen based on a comparative evaluation of two motion artifact detection approaches (see supplementary material [5]).

#### 2) Blood Volume Pulse (BVP)

Motion-related artefacts identified from the acceleration recordings were removed from the BVP signal, and the resulting gaps were interpolated prior to heartbeat detection. Heart rate (HR) and heart rate variability (HRV) metrics were then calculated from the detected, physiologically plausible beat intervals. Since BVP recordings are particularly susceptible to motion artefacts, six pre-processing pipelines with different artefact-removal and signal-processing strategies were compared to select the most appropriate one for the study data. All pipelines yielded comparable model performance [5]. This indicates that the choice of pre-processing pipeline for our physiological BVP data will not have a major impact on the subsequent statistical analyses and interpretations of results. We selected the pre-processing pipeline that retained the most working tasks. For each working task, the following cardiovascular metrics were calculated:

- **mean_HR**: mean HR (bpm)
- **HRV_SDNN**: standard deviation of inter-beat intervals (IBIs), reflecting overall heart rate variability (both sympathetic and parasympathetic influences)
- **HRV_RMSSD**: root mean square of successive differences between adjacent IBIs, reflecting parasympathetic activity
- **HRV_pNN50**: percentage of adjacent IBI pairs that differ more than 50ms, reflecting short-term HRV and parasympathetic regulation

Cardiovascular metrics, including HR and HRV, have been associated with psychological stress and mental workload in previous research [12, 21].

#### 3) Electrodermal Activity (EDA)

High-motion segments were also filtered out of the EDA signals and the gaps were interpolated. The signal was then decomposed into its tonic and phasic components and skin conductance responses (SCRs) were detected in the phasic component of the signal. For each working task, the following metrics were extracted:

- **mean_SCL**: mean skin conductance level (SCL), i.e. the mean of the tonic signal component
- **mean_phasic**: mean of the phasic signal component, reflecting the average magnitude of skin conductance fluctuations
- **mean_cleaned_EDA**: mean of the cleaned EDA signal containing both the tonic and phasic signal components
- **SCRs_norm_duration**: number of SCRs normalized by the task duration in seconds, representing the rate of phasic electrodermal responses

Electrodermal metrics reflect the activity of the sympathetic nervous system and have been linked to physiological arousal and mental workload [13, 27].

#### 4) Skin Temperature (TEMP)

Motion artefacts were also removed from the TEMP signal and implausible values were removed before calculating the following metrics:

- **mean_TEMP**: mean skin temperature
- **std_TEMP**: standard deviation of the skin temperature
- **range_TEMP**: range of skin temperature values

Skin temperature may provide complementary information on peripheral physiological responses that may accompany changes in stress, emotional arousal, and cognitive workload [7, 34].

#### 5) Summary of the Data

TABLE I. summarizes the physiological dataset used for the statistical analyses after pre-processing: After physiological data cleaning, each participant's working day contained at least 83.23% (BVP), 74.6% (EDA), and 73.09% (TEMP) of its original data. All participants were included in the EDA and TEMP analyses, whereas two participants were excluded from the BVP data analyses due to insufficient data quality.

Across the dataset, more working tasks were excluded during pre-processing for BVP than for EDA and TEMP.

Fewer tasks can therefore be analyzed based on BVP data. Among the retained tasks, development-heavy activities constitute the largest category, followed by collaboration-heavy and other tasks. Average task durations were computed for each combination of task category and physiological modality. Across all combinations, durations are comparable, ranging from 49.9 to 51.38 minutes.

TABLE I. OVERVIEW OF THE PHYSIOLOGICAL DATASET AFTER PRE-PROCESSING USED FOR STATISTICAL ANALYSES

| | BVP | EDA | TEMP |
|---|---|---|---|
| Min % of original data in a working day | 83.23% | 74.6% | 73.09% |
| Number of participants included | 19 | 21 | 21 |
| Number of working tasks included | 263 (out of 382) | 371 (out of 382) | 371 (out of 382) |
| *Development-heavy tasks* | *121* | *158* | *159* |
| *Collaboration-heavy tasks* | *76* | *108* | *106* |
| *Other tasks* | *66* | *105* | *106* |

## F. Data Analysis – Linear Mixed-Effects Model

To investigate whether physiological activity, GenAI use, task category and task duration were associated with developers' perceived cognitive load, we fitted a series of linear mixed-effects models. The lme4 and lmerTest packages in R were used for the analyses. All models included participant as a random intercept to account for repeated measures within participants and to capture between-participant differences in cognitive load ratings.

### 1) Baseline Model

First, we fit a baseline model to assess the association between GenAI use, task category, task duration, and perceived cognitive load in the physiologically filtered dataset. The model included GenAI use (yes/no), task category (development-heavy, collaboration-heavy, or other), and task duration in minutes as fixed effects and participant as a random intercept:

$$\text{Cognitive load} \sim \text{GenAI use} + \text{task category} + \text{task duration} + (1|\text{participant}) \quad (1)$$

This baseline model served two purposes: First, it was used to assess the contribution of the task characteristics to perceived cognitive load. Second, it served as the baseline model for evaluating whether physiological metrics provided additional explanatory value beyond GenAI use, task category, task duration, and participant-specific differences.

### 2) Individual Physiological Metrics

We then investigated whether individual physiological metrics provided additional information about perceived cognitive load beyond the work context baseline (1). Separate models were fit for each physiological metric (four HR/HRV, four EDA, and three skin temperature metrics) (2) and were then compared to the baseline model (1).

$$\text{Cognitive load} \sim \text{physiological metric} + \text{GenAI use} + \text{task category} + \text{task duration} + (1|\text{participant}) \quad (2)$$

Before model fitting, physiological metrics with non-normal distributions (e.g., RMSSD) were log-transformed, and all metrics were z-standardized per participant.

### 3) Multimodal Physiological Analysis

In addition to the individual-metric analyses, we examined whether physiological characteristics jointly provided information about perceived cognitive load. Before constructing the multimodal model, we examined the correlations among the physiological metrics to avoid redundant measurements influencing the analyses. Metrics correlating above $|r|=0.90$ were removed (HRV_SDNN, mean_cleaned_EDA, range_TEMP).

Principal component analysis (PCA) was then conducted separately for the BVP and EDA metrics. Parallel analysis supported retaining one BVP component and one EDA component. The BVP component broadly represented higher heart rate in combination with lower HRV, whereas the EDA component primarily represented higher tonic activity and SCR frequency.

According to the correlation analysis, mean_TEMP and std_TEMP were not correlated and therefore represented distinct aspects of the skin temperature signal. They were therefore retained as separate predictors in the multimodal model rather than combined into a single component. The final multimodal model (4) therefore included the BVP principal component ($PC_{BVP}$), the EDA principal component ($PC_{EDA}$), mean_TEMP, and std_TEMP as fixed effects, together with the work context covariates from the baseline model (1).

$$\text{Cognitive load} \sim \text{GenAI use} + \text{task category} + \text{task duration} + PC_{BVP} + PC_{EDA} + \text{mean_TEMP} + \text{std_TEMP} + (1|\text{participant}) \quad (4)$$

### 4) Model Comparisons and P-Value Adjustments

Likelihood-ratio tests (LRTs) were used to compare nested models. Benjamini–Hochberg multiple comparison correction was applied to the relevant families of hypothesis tests. Tukey adjustment was used for pairwise post hoc comparisons among task categories in the baseline model (1). Standard diagnostic checks were used to assess model assumptions and fit.

# IV. RESULTS

## A. Work Context Factors Associated with Perceived Cognitive Load During Everyday Working Tasks

Using the behavioral baseline model (1) defined in Section III-F, we first examined whether GenAI use, task category, and task duration were associated with developers' perceived cognitive load during everyday working tasks.

GenAI use, task category, and task duration each significantly contributed to explaining developers' perceived cognitive load (TABLE II. ). GenAI use during working tasks was associated with approximately a one-point higher cognitive load rating on the seven-point scale compared to tasks not involving GenAI use (estimate=0.996, $\chi^2(1)=23.80$, $p<0.001$, $\Delta R^2=0.047$) (Fig. 3, left). Furthermore, perceived cognitive load increased by 0.007 points for every additional minute of task duration (estimate=0.007, $\chi^2(1)=15.21$, $p<0.001$, $\Delta R^2=0.038$). The task category added the largest explanatory value to the model and therefore to explaining perceived cognitive load ($\chi^2(2)$=44.13, p<0.001, $\Delta R^2$=0.091). Post hoc pairwise comparisons (Tukey-adjusted) showed that development-heavy tasks were associated with significantly higher perceived cognitive load than collaboration-heavy tasks (estimate=0.591, p=0.0153) and other activities (estimate=1.437, p<0.0001) (Fig. 3, right). Furthermore, collaboration-heavy tasks were associated with significantly higher perceived cognitive load than other activities (estimate=0.845, p<0.0001) (Fig. 3, right). Overall, the behavioral model accounted for 32.4% of the variance in perceived cognitive load through the three fixed effects and 45.9% when participant-specific differences were included.

In sum, perceived cognitive load during everyday working tasks was associated with the work context: it was higher during GenAI use, increased with task duration, and differed substantially across task categories, with development-heavy tasks showing the highest cognitive load.

TABLE II. LIKELIHOOD-RATIO TEST RESULTS FOR THE BEHAVIORAL MODEL

| Behavioral factor | Estimate | $\chi^2$ | df | p | $\Delta R^2$ |
|---|---|---|---|---|---|
| GenAI use | 0.996 | 23.80 | 1 | < 0.001 | 0.047 |
| Task category | - | 44.14 | 2 | < 0.001 | 0.091 |
| Task duration | 0.007 | 15.21 | 1 | < 0.001 | 0.038 |

[a] $\Delta R^2$ denotes the increase in marginal $R^2$ when the respective behavioral factor is included in the model. The full model achieved a marginal $R^2$ of 0.324 and a conditional $R^2$ of 0.459. Marginal $R^2$ represents variance explained by fixed effects and conditional $R^2$ represents variance accounted for by both fixed and random effects.

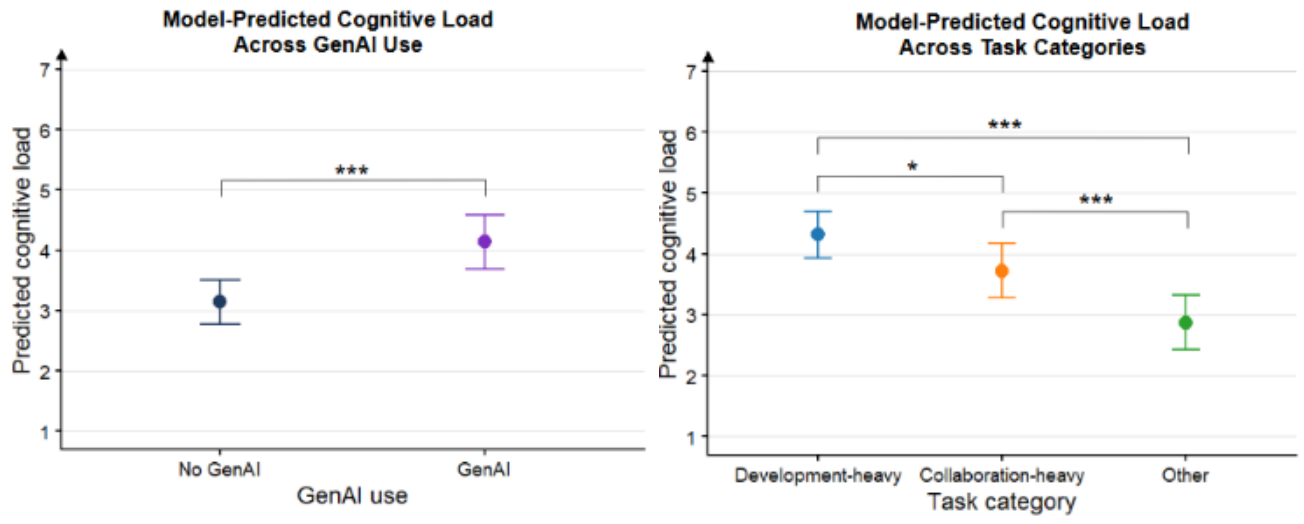

Fig. 3. Model-predicted perceived cognitive load based on the contextual mixed-effect model across working tasks without or with GenAI use (left) and across the three task categories (right). Points show estimated marginal means; error bars indicate 95% confidence intervals. Asterisks denote statistical significance (* p≤0.05, *** p≤0.001).

## B. RQ2: Limited Additional Information from Wearable-derived Physiological Characteristics

To evaluate whether wearable-derived physiological characteristics provided additional information about developers' perceived cognitive load beyond the contextual baseline model (1), we first examined individual physiological metrics and then evaluated whether considering multiple physiological modalities together provided additional information.

### *1) Individual Physiological Metrics*

Via the physiological model structure (2) defined in Section III-F, each of the eleven physiological metrics was added separately to the contextual baseline model and evaluated against the baseline model using a likelihood-ratio test.

Across all eleven physiological metrics, skin temperature standard deviation (std_TEMP) was the only metric that significantly improved model fit after correction for multiple comparisons (TABLE III. ). Lower skin temperature variability was associated with higher perceived cognitive load (estimate=-0.224, $\chi^2(1)$=8.43, $p_{BH}$=0.041). However, the additional explanatory contribution was small ($\Delta R^2$=0.014). Skin temperature range (range_TEMP) showed a similar negative association, but did not remain statistically significant after Benjamini-Hochberg correction ($p_{BH}$=0.105). Mean skin temperature (mean_TEMP) showed a small positive estimate but did not substantially improve model fit.

As a sensitivity analysis, we repeated the analyses using cumulative link mixed models to account for the ordinal nature of the cognitive load ratings. The results were consistent with the linear mixed-effects models: only skin temperature standard deviation significantly improved model fit after multiple comparison correction (estimate=-0.314, $\chi^2(1)$=8.56, $p_{BH}$=0.038), while the remaining individual physiological metrics did not improve model fit beyond the work context factors.

As an additional sensitivity check for the effect of the rating scale granularity, we grouped the seven-point cognitive load Likert scale into three broader ordinal categories - low (1-2), medium (3-5), and high (6-7) - and repeated the analyses with the cumulative link mixed models. The overall patterns of the results remained mainly unchanged, but with skin temperature standard deviation no longer remaining statistically significant after multiple comparison correction (estimate=-0.331, $\chi^2(1)$=6.77, $p_{BH}$=0.102).

In sum, after accounting for GenAI use, task category and task duration and participant-specific differences, skin temperature standard deviation was the only individual physiological metric that improved the model's ability to account for variation in cognitive-load ratings. Sensitivity analyses showed consistent directions of the associations between physiological metrics and cognitive load ratings.

TABLE III. LIKELIHOOD-RATIO TEST RESULTS FOR INDIVIDUAL PHYSIOLOGICAL METRICS

| Physiological signal | Metric | Estimate | $\chi^2$ | df | p | $P_{BH}$ | $\Delta R^2$ |
|---|---|---|---|---|---|---|---|
| Cardiovascular activity | Mean_HR | -0.010 | 0.02 | 1 | 0.900 | 0.900 | <0.001 |
| | HRV_lnRMSSD | -0.052 | 0.41 | 1 | 0.521 | 0.637 | <0.001 |
| | HRV_SDNN | -0.084 | 1.07 | 1 | 0.301 | 0.552 | 0.002 |
| | HRV_pNN50 | 0.057 | 0.51 | 1 | 0.477 | 0.637 | 0.001 |
| Electrodermal activity | Mean_SCL | 0.079 | 1.29 | 1 | 0.256 | 0.552 | 0.002 |
| | Mean_phasic | 0.037 | 0.28 | 1 | 0.594 | 0.654 | <0.001 |
| | Mean_cleaned_EDA | 0.079 | 1.29 | 1 | 0.255 | 0.552 | 0.002 |
| | SCR_norm_duration | 0.090 | 1.58 | 1 | 0.208 | 0.552 | 0.002 |
| Skin temperature | Mean_TEMP | 0.058 | 0.69 | 1 | 0.412 | 0.637 | <0.001 |
| | Std_TEMP | -0.224 | 8.43 | 1 | **0.004** | **0.041** | 0.014 |
| | Range_TEMP | -0.181 | 5.49 | 1 | **0.019** | 0.105 | 0.009 |

[b] Each physiological model added one physiological metric to the contextual baseline model and was compared with that baseline using a likelihood ratio test(LRT). The baseline and physiological models in each comparison were fitted to the same observations. Physiological metrics were z-standardized before model fitting. Consequently, the estimate of each physiological metric represents the expected change

in cognitive load ratings associated with a one standard deviation increase in the physiological metric, after accounting for GenAI use, task category, task duration, and participant-specific differences. $\Delta R^2$ denotes the increase in marginal $R^2$ when adding the physiological metric to the baseline model. LRT p-values were adjusted with the Benjamini–Hochberg (BH) method to account for multiple comparisons across the eleven physiological metrics.

TABLE IV. LIKELIHOOD-RATIO TEST RESULTS FOR THE MULTIMODAL PHYSIOLOGICAL ANALYSIS

| Model comparison | Estimate | $\chi^2$ | df | p | $p_{BH}$ | $\Delta R^2$ |
|---|---|---|---|---|---|---|
| Complete multimodal model vs contextual baseline model | - | 9.42 | 4 | 0.052 | - | 0.019 |
| *Single modality contribution* | | | | | | |
| Cardiovascular component ($PC_{BVP}$) | -0.011 | 0.04 | 1 | 0.844 | 0.844 | <0.001 |
| Electrodermal component ($PC_{EDA}$) | 0.075 | 1.31 | 1 | 0.253 | 0.379 | 0.002 |
| Skin temperature metrics (mean_TEMP + std_TEMP) | - | 8.87 | 2 | **0.012** | **0.035** | 0.018 |
| *TEMP follow-up comparison* | | | | | | |
| Mean_TEMP | -0.025 | 0.09 | 1 | 0.762 | 0.762 | <0.001 |
| Std_TEMP | -0.251 | 8.83 | 1 | **0.003** | **0.006** | 0.018 |

[c.] The complete multimodal model included the cardiovascular principal component ($PC_{BVP}$), the electrodermal principal component ($PC_{EDA}$), mean skin temperature (mean_TEMP), and skin-temperature standard deviation (std_TEMP). Estimates are reported only for single-predictor tests; no single estimate is available for the combined physiological block or the two-predictor temperature block. The single modality likelihood ratio tests compared the complete multimodal model with models excluding the respective modality. Their p-values were adjusted across the three modalities using the Benjamini–Hochberg procedure. Temperature follow-up p-values were adjusted separately across mean_TEMP and std_TEMP. $\Delta R^2$ denotes the increase in marginal $R^2$ associated with adding the tested block or predictor to the corresponding reduced model.

### 2) Multimodal Physiological Characteristics

To evaluate whether physiological information became more informative when multiple modalities were considered jointly, we extended the contextual baseline model by adding a cardiovascular principal component, an electrodermal-activity principal component, mean skin temperature, and skin-temperature standard deviation (Section III-F, (3)). The first cardiovascular component explained 68.8% of the variance across mean heart rate, lnRMSSD, and pNN50 and broadly represented higher heart rate combined with lower heart-rate variability. The first EDA component explained 45.7% of the variance and primarily captured higher SCL and SCR frequency with a weaker contribution of lower mean phasic activity.

Comparing the contextual baseline model with the model that additionally contained all four physiological predictors ($PC_{BVP}$, $PC_{EDA}$, mean_TEMP and std_TEMP) showed a nearly statistically significant contribution ($\chi^2(4)=9.42$, $p=0.052$) (TABLE IV. ). The combined physiological parameters increased the model's ability to account for additional variation in the cognitive load ratings beyond GenAI use, task category, task duration, and participant-specific differences. However, the additional explanatory contribution was small ($\Delta R^2=0.019$).

Leave-one-modality-out model comparisons indicated that most additional information was contained in the skin temperature (TABLE IV. ). Removing mean_TEMP and std_TEMP significantly reduced model fit ($\chi^2(2)=8.87$, $p_{BH}=0.035$, $\Delta R^2=0.018$). Neither $PC_{BVP}$ nor $PC_{EDA}$ significantly improved model fit ($PC_{BVP}$: estimate=-0.011, $\chi^2(1)=0.04$, $p_{BH}=0.844$, $\Delta R^2<0.001$; $PC_{EDA}$: estimate=0.075, $\chi^2(1)=1.31$, $p_{BH}=0.379$, $\Delta R^2=0.002$).

Follow-up comparisons of the contribution of the individual skin temperature factors mean_TEMP and std_TEMP indicated that the significant skin temperature result was driven by std_TEMP. Lower std_TEMP was associated with higher cognitive load ratings (estimate=-0.251, $\chi^2(1)=8.83$, $p_{BH}=0.006$, $\Delta R^2=0.018$) (TABLE IV. ). Accordingly, a one-standard-deviation increase in skin-temperature variability was associated with an approximately 0.25-point decrease in perceived cognitive-load ratings after accounting for the contextual factors, participant-specific differences, and the remaining physiological predictors. Mean_TEMP provided no detectable further information.

As a sensitivity analysis, we repeated the multimodal comparison using cumulative link mixed models to account for the ordinal nature of the seven-point cognitive load ratings scale. The physiological model containing all four physiological predictors also significantly improved model fit over the contextual baseline model ($\chi^2(4)=10.89$, $p=0.028$). The two skin temperature metrics together also contribute significantly to the model fit ($\chi^2(2)=10.46$, $p_{BH}=0.016$), whereas $PC_{BVP}$ and $PC_{EDA}$ did not. As in the linear mixed-effects analysis, std_TEMP contributed significantly (estimate=-0.396, $\chi^2(1)=10.42$, $p_{BH}=0.002$), while mean_TEMP remained non-significant.

In sum, the multimodal analysis provided evidence that wearable-derived physiological information from BVP, EDA and skin temperature contained marginal but some additional information about perceived cognitive load beyond GenAI use and task context. The most informative metric was the standard deviation of skin temperature, while the cardiovascular and EDA components made no major contribution. This finding is consistent with the individual physiological analyses (RQ2), in which skin-temperature standard deviation was likewise the only metric that significantly improved model fit after correction for multiple comparisons.

## V. DISCUSSION

### A. GenAI Use, Task Category, Task Duration are Associated with Perceived Cognitive Load

The results show that GenAI use was associated with higher perceived cognitive load during everyday software development tasks, even after accounting for task category, task duration, and participant-specific differences. Task category and task duration were also associated with cognitive load, indicating that GenAI use should be interpreted within the broader behavioral work context.

**GenAI Use**

Tasks involving GenAI use were associated with nearly one point higher perceived cognitive load rating than tasks without GenAI use. This finding should not be interpreted as evidence that GenAI use generally makes software development more difficult, nor does it contradict prior findings that developers perceive GenAI as useful [11]. Rather, it suggests that GenAI-supported work during everyday software development is not necessarily experienced as cognitively less demanding. GenAI interaction may introduce additional cognitive demands, such as formulating prompts, evaluating generated outputs for correctness and usefulness, adapting suggestions, and integrating generated content into an existing codebase.

This interpretation is consistent with previous findings from the same broader study [8]: In the controlled task setting, GitHub Copilot supported work efficiency and reduced perceived workload when the interaction type suited the task. Using multiple interaction types may have reflected uncertainty about how best to interact with the assistant, which can require additional cognitive effort [8]. During everyday work, GenAI-supported tasks were associated with both higher perceived cognitive load and higher productivity [8]. Questionnaire results of the study further showed that developers described work efficiency and productivity as positive aspects of GenAI use, but also reported challenges related to prompt sensitivity, output inaccuracies, and the need to validate generated output [10].

Our task-level findings on perceived cognitive load differ from more general assessments reported in prior work. Vella and Blincoe found that developers generally perceived AI as beneficial for cognitive load [33], and the developers in this present study similarly tended to agree that working with AI is less mental effort compared to working without it [10]. In contrast, the task-level analysis associated GenAI-supported tasks with higher perceived cognitive load. This divergence may reflect discrepancies between global retrospective evaluations, during which experiences are aggregated and can be influenced by more salient or recent events, and ratings of specific activities [23]. The findings are not exclusive: developers may perceive AI as reducing cognitive load overall, while individual AI-supported tasks may still be experienced as more cognitively demanding than comparable tasks completed without AI.

Together, the present and previous findings [8, 10, 33] suggest that GenAI use does not simply reduce developers' cognitive effort. Instead, perceived cognitive workload during GenAI-supported development could reflect an interplay between GenAI use, interaction type and intensity, output quality, and task characteristics [8, 10]. These factors may determine how much effort GenAI use takes or saves. AI-supported work may shift effort from producing solutions manually to supervising, evaluating, and integrating AI-generated output. Vella and Blincoe observed that developers reported spending less time on direct code creation while describing increased effort in directing, reviewing, and correcting AI output [33]. This cognitive offloading to AI during implementation can come at a cost described by Alakmeh et al. as comprehension debt [1] - the growing gap between what developers can produce and what they understand. This broader concern is also reflected in Storey's Triple Debt theory [28], which highlights how increasing reliance on AI-generated software may create longer-term costs for system understanding and maintainability. While the present study does not measure these forms of debt or cognitive offloading directly, its task-level findings are consistent with prior evidence that AI-supported development can remain cognitively demanding despite reduced manual code production.

**Task Category**

Task category showed the largest explanatory value for perceived cognitive load among the behavioral factors. Development-heavy tasks were associated with the highest perceived cognitive load, followed by collaboration-heavy tasks and other activities. Consistent with previous findings [7, 10] the present analysis shows that task category remains strongly associated with perceived cognitive load even after accounting for GenAI use, task duration, and participant-specific differences.

The higher perceived cognitive load during development-heavy tasks may reflect the developers' need to understand requirements, make implementation decisions, integrate changes into the existing codebase, review and test the code etc. The finding that collaboration-heavy tasks were associated with higher perceived cognitive load than other activities further shows that cognitive load in software engineering is not limited to coding or implementation work. It may also arise from coordination and communication activities. These results show that task context is a central factor for understanding perceived cognitive load during everyday software development.

**Task Duration**

Task duration was also positively associated with perceived cognitive load, although the estimated increase per minute was small. This result is consistent with previous work [7], which found a positive correlation between task duration and cognitive load ratings but no difference across task categories. The present analysis extends this finding by showing that this association remains even after accounting for task category, GenAI use, and participant-specific differences. Thus, longer tasks may reflect sustained attention or more complex work, which contributes to perceived cognitive load during everyday software development tasks.

Overall, GenAI use, task category, and task duration each accounted for a statistically significant share of variation in perceived cognitive load. These findings establish the contextual baseline for RQ2, which examines whether wearable-derived physiological characteristics provide additional information about perceived cognitive load beyond GenAI use and task context.

### *B. Wearable-Derived Physiological Characteristics Provide Limited but Detectable Additional Information Beyond Work Context*

**Individual Physiological Metrics**

Most individual physiological metrics did not provide detectable additional information about variations in perceived cognitive load ratings after GenAI use, task category, task duration, and participant-specific differences were accounted for. Skin-temperature standard deviation was the only exception: lower within-task temperature variability was associated with higher perceived cognitive load.

Ordinal sensitivity analyses yielded consistent results, whereas collapsing the seven-point cognitive load scale into three broader categories removed the significance of std_TEMP, suggesting information loss due to broader groups. Therefore, the results of the linear mixed-effects models with a seven-point cognitive load rating scale are discussed here.

The absence of detectable associations across most individual metrics illustrates a key challenge in transferring physiological cognitive load measurements from controlled laboratory settings to everyday, real-world software engineering settings. Controlled studies using n-back, Stroop, and arithmetic tasks have demonstrated physiological responses to experimentally induced cognitive load levels [3, 15, 25, 29]. In real-world software development, however, cognitive load differences are more subtle and physiological signals are simultaneously influenced by surrounding work conditions, movement, emotional states, device fit and other uncontrolled factors. These influences may partly explain why most individual metrics provided no detectable information beyond work context in the present study.

Against this background, the persistence of a temperature-variability association with perceived cognitive load is noteworthy: some task-level physiological information remained detectable despite the uncontrolled field conditions. Higher perceived cognitive load was associated with more stable skin temperature during working tasks, while mean skin temperature was not significantly associated. Reduced variability could reflect sustained peripheral physiological regulation, stable posture and environmental conditions, or other processes that covary with cognitively demanding work.

These results relate to, but do not replicate, our findings in previous work, in which we found that mean skin temperature was positively correlated with perceived cognitive load [7]. Whereas the previous study found an association with mean skin temperature, the present study identifies an association with within-task temperature variability. Predictive analyses of cognitive load with XGBoost using a dataset from the previous real-world study [7] found that working task category, EDA and skin temperature were relevant in feature selection analyses [31]. This aligns with the present findings that work context accounts for a substantial share of variation in perceived cognitive load ratings and that skin temperature could be a direct or indirect indicative feature for cognitive load. Taken together, the studies suggest that the dynamics of skin temperature could be worth further investigation in real-world cognitive load research.

**Multimodal Physiological Characteristics**

We next examined whether wearable-derived physiological information became more informative when considered jointly in a multimodal and context-aware analysis. Cardiovascular activity, EDA, and skin temperature characteristics were combined into one physiological model and compared to the contextual baseline model. Adding multimodal physiological information to the contextual baseline model provided an almost statistically significant, though small, amount of additional information about cognitive load variation. This result is noteworthy since despite substantial uncontrolled variation in everyday work environment, wearable-derived physiology contained a small but detectable amount of task-level information associated with perceived cognitive load ratings after GenAI use, task context and participant-specific differences were considered.

The leave-one-modality-out analyses clarified the source of this information. Removing the two skin temperature characteristics significantly reduced model fit, whereas removing the cardiovascular or EDA component did not. Follow-up comparisons showed that the temperature result was again attributable to skin temperature standard deviation rather than mean temperature. Thus, the multimodal analysis does not demonstrate a broadly informative physiological pattern distributed across all modalities. Instead, the findings support the individual-metric result that the temperature variability association remained detectable even after accounting for cardiovascular activity, EDA, and mean temperature. The findings indicate conditional robustness of skin-temperature variability rather than for multimodal synergy among all investigated signals. Small or shared physiological effects may still be present, but the current models did not isolate them.

Complementary predictive work on our previous field-study dataset [7] found that multimodal physiological and task-context information can support cognitive load prediction, particularly in personalized models such as XGBoost [31]. Predictive models may exploit nonlinear relations, interactions, and shared information across signals, whereas the present linear mixed-effects analysis assesses interpretable, adjusted associations and the unique contribution of physiological characteristics beyond work context. The two approaches are therefore complementary: the present results identify temperature variability as the clearest adjusted physiological association, while the predictive findings motivate continued investigation of multimodal and personalized representations.

All in all, the individual and multimodal analyses provide modest evidence that wearable-derived physiological data contains additional information about perceived cognitive load beyond GenAI use and work context. The clearest association was with skin-temperature variability, while cardiovascular and electrodermal characteristics provided no detectable additional contribution. The small physiological contribution and its concentration in one characteristic do not support treating wearable physiology as a context-independent measure of cognitive load.

### *C. Implications for Software Engineering Research and Practice*

Our findings have implications for how organizations evaluate GenAI-supported software development. Productivity gains, code output quality, and tool adoption metrics should not be interpreted as evidence that GenAI-supported work is perceived as cognitively less demanding. GenAI-supported tasks were associated with higher perceived cognitive load ratings after accounting for task category and duration. Evaluations of GenAI tools and developers' interaction with them could therefore complement productivity and output quality measures with task-level developer experience measures that also account for the work contexts in which GenAI is used. This can help distinguish cognitive demands associated with the underlying engineering task from demands related to prompting, validating, correcting, and integrating AI-generated output.

The results also provide practical guidance for wearable-based developer experience studies in real-world settings. Individual physiological metrics should not be interpreted as direct measures of perceived cognitive load but analyzed together with work context information, such as task type, duration, and GenAI use. The multimodal results support combining physiological and contextual information, but not replacing task-level self-reports or work context measures with wearable sensing.

At the present stage, wearable-derived cognitive load information is better suited to aggregate evaluations of tools, workflows, and recurring work conditions than for individual performance assessment or automated workplace decisions. The practical value of wearables lies in their potential to help identify whether tools and work designs systematically change or create novel and potentially avoidable cognitive demands, as a continuous and objective complement to subjective self-reports and behavioral information.

For development leaders, these findings suggest embedding GenAI evaluation in a broader measurement approach rather than treating it solely on productivity, adoption, usage and delivery metrics that do not capture developers' cognitive experience. Assessing GenAI impact across recurring work contexts, such as development-heavy implementation tasks, review activities and collaboration-heavy work, could help identify where workflow design, guidance, training and tool integration could reduce avoidable cognitive load and support more cognitively sustainable AI-supported work.

## VI. Threats to Validity and Future Work

### A. Construct Validity

A few aspects may impact the construct validity of this study. First, cognitive load was measured via subjective ratings, which capture developers' perceived cognitive load and may vary with individuals'scale use and self-assessment. Sensitivity analyses using ordinal scale models and a three-category regrouping yielded broadly comparable results, although the strength of evidence depended partly on the original scale granularity. Nevertheless, self-reported cognitive load remains only an indirect measure of actual cognitive processes.

Secondly, physiological measures are not specific to cognitive load and are also affected by movement, affect, environmental conditions and other internal and external processes. The significant temperature variability result therefore demonstrates an association with perceived cognitive load ratings, rather than a direct measure of cognitive load.

Thirdly, GenAI use was measured as a binary task property. This binary measure aligns with the study's research question but does not distinguish interaction type, interaction intensity, output quality, or the extent of validation and correction, all of which could help explain heterogeneity in perceived cognitive load among GenAI-supported tasks.

Finally, working tasks were documented at the granularity chosen by developers to minimize disruption and preserve ecological validity. This limited the amount and temporal precision of data available for linking working tasks, cognitive load ratings and physiological signals.

### B. Internal Validity

Real-world wearable data are affected by contextual and emotional factors, movement, device fit, and other uncontrolled influences. Although we pre-processed the physiological signals to remove strong motion artefacts and implausible values, we could not eliminate all non-cognitive sources of variation in the everyday setting.

Additional simulation-based sensitivity analyses suggested that the available sample size and repeated-measures structure provided adequate sensitivity to detect moderate individual physiological associations but lower sensitivity for small associations and distributed multimodal effects. The non-significant physiological findings should therefore not be interpreted as evidence of no association, but rather that smaller physiological associations may have remained undetected.

### C. External Validity

The study was conducted with professional software developers in their everyday work environment, which strengthens the ecological relevance of our findings. However, a larger participant group, a longer study period, and data collection across multiple organizations could strengthen their transferability.

Because GenAI tools and interaction paradigms are evolving rapidly, especially with the increasing use of agentic systems, replication across organizations, roles, project types, working tasks, wearable platforms, and newer forms of AI-supported development is needed before generalizing the observed associations broadly.

### D. Future Work

Future work could build on this study design and collect larger real-world datasets across multiple organizations to improve statistical power, generalizability, and to deduce if cognitive load arises from AI interaction or from the actual working task. More fine-grained GenAI interaction measures, including prompting characteristics, in-code completion, agentic delegation, validation effort, interaction intensity, and output quality, could clarify in more detail which forms of AI-supported work introduce or reduce cognitive demands. More work context features, such as task complexity, interruptions, codebase familiarity etc. could also add to a more fine-grained analysis.

Extended datasets could also be used to try multimodal predictive approaches. As measurements of cognitive load via wearables in real-world settings become more reliable, such information from physiological data could support code and tool evaluation, task scheduling, and long-term developers' well-being in their everyday work contexts.

Future studies should examine GenAI-supported development as part of a broader developer experience enterprise ecosystem, which includes IDE assistants, chat-based assistants, agentic systems, internal documentation systems, and platform engineering services. This would allow organizations to understand how these tools reshape developer workflows, coordination patterns and sustainable productivity.

## VII. Conclusion

This study examined how GenAI use and task context relate to perceived cognitive load during real-world software development and whether wearable-derived physiological characteristics provide additional information beyond these factors. GenAI use, task category, and task duration accounted for substantial variation in perceived cognitive load. Wearable-derived physiology provided only limited additional information: across individual and multimodal analyses, skin-temperature variability showed the clearest association, while cardiovascular and electrodermal characteristics provided no significant additional contribution. Overall, these findings suggest that perceived cognitive load in ecological software engineering settings is more strongly associated with GenAI use and work context than with wearable-derived physiological measures. Wearable data should therefore be interpreted in relation to the developers' concrete work context.

## REFERENCES

[1] Tarek Alakmeh, Norman Anderson, Victoria Jackson, Guilherme Vaz Pereira, Umit Akirmak, Anthony Estey, Rafael Prikladnicki, André van der Hoek, Margaret-Anne Storey, and Thomas Fritz. 2026. Grasping AI Reliance in Program Comprehension and Coding through the AIRELI Persona Taxonomy. In *Proceedings of the 2026 34th IEEE/ACM International Conference on Program Comprehension*, April 12, 2026. ACM, New York, NY, USA, 123–134. https://doi.org/10.1145/3794763.3794804

[2] Christoph Anders, Sidratul Moontaha, Samik Real, and Bert Arnrich. 2024. Unobtrusive measurement of cognitive load and physiological signals in uncontrolled environments. *Sci. Data* 11, 1 (September 2024), 1000. https://doi.org/10.1038/s41597-024-03738-7

[3] Paul Ayres, Joy Yeonjoo Lee, Fred Paas, and Jeroen J.G. van Merriënboer. 2021. The Validity of Physiological Measures to Identify Differences in Intrinsic Cognitive Load. *Frontiers in Psychology 12*. https://doi.org/10.3389/fpsyg.2021.702538

[4] Vadim Borisov, Enkelejda Kasneci, and Gjergji Kasneci. 2021. Robust cognitive load detection from wrist-band sensors. *Computers in Human Behavior Reports* 4, (August 2021), 100116. https://doi.org/10.1016/j.chbr.2021.100116

[5] Charlotte Brandebusemeyer, Daniela Gasser, Tobias Schimmer, and Bert Arnrich. 2026. Supplementary Material for Beyond Productivity: Measuring Developers' Cognitive Load During GenAI-Supported Software Development. *Zenodo*. https://doi.org/10.5281/zenodo.22692764

[6] Charlotte Brandebusemeyer, Fabian Georgi, and Bert Arnrich. 2026. Reliability Assessment of Wearable Technologies for Physiological Measurements: An Evaluation of Shimmer3 GSR+, Empatica E4, EmbracePlus, and Pixel Watch 2 Across Cognitive, Affective and Physical Activity Tasks. *Sensors* 26, 14 (July 2026), 4376. https://doi.org/10.3390/s26144376

[7] Charlotte Brandebusemeyer, Tobias Schimmer, and Bert Arnrich. 2025. Wearables to Measure Developer Experience at Work. In *2025 IEEE/ACM 47th International Conference on Software Engineering: Software Engineering in Practice (ICSE-SEIP)*, April 27, 2025. IEEE, 23–33. https://doi.org/10.1109/ICSE-SEIP66354.2025.00008

[8] Charlotte Brandebusemeyer, Tobias Schimmer, and Bert Arnrich. 2026. Developers' Experience with Generative AI - First Insights from an Empirical Mixed-Methods Field Study. In *Proceedings of the IEEE/ACM 48th International Conference on Software Engineering: Software Engineering in Practice*, April 12, 2026. ACM, New York, NY, USA, 259–268. https://doi.org/10.1145/3786583.3786870

[9] Charlotte Brandebusemeyer, Fabian Stolp, and Bert Arnrich. 2026. Cognitive Load in Programming - An Interdisciplinary Systematic Mapping Review. *SSRN, Manuscript submitted for review*. https://doi.org/10.2139/ssrn.7477619

[10] Charlotte Brandebusemeyer, Kerim Zunic, Thomas Zimmermann, Tobias Schimmer, and Bert Arnrich. 2026. Developers' Experience with Generative AI Beyond Productivity Assessment - Insights from an Empirical Mixed-Methods Field Study. *ArXiv* (2026). Retrieved July 13, 2026 from https://arxiv.org/abs/2607.02337

[11] Jenna Butler, Jina Suh, Sankeerti Haniyur, and Constance Hadley. 2025. Dear Diary: A Randomized Controlled Trial of Generative AI Coding Tools in the Workplace. In *2025 IEEE/ACM 47th International Conference on Software Engineering: Software Engineering in Practice (ICSE-SEIP)*, April 27, 2025. IEEE, 319–329. https://doi.org/10.1109/ICSE-SEIP66354.2025.00034

[12] Rebecca L. Charles and Jim Nixon. 2019. Measuring mental workload using physiological measures: A systematic review. *Appl. Ergon.* 74, (January 2019), 221–232. https://doi.org/10.1016/j.apergo.2018.08.028

[13] Hugo D. Critchley. 2002. Electrodermal Responses: What Happens in the Brain. *The Neuroscientist* 8, 2 (April 2002), 132–142. https://doi.org/10.1177/107385840200800209

[14] Kevin Zheyuan Cui, Mert Demirer, Sonia Jaffe, Leon Musolff, Sida Peng, and Tobias Salz. 2026. The Effects of Generative AI on High-Skilled Work: Evidence from Three Field Experiments with Software Developers. *Manage. Sci.* (February 2026). https://doi.org/10.1287/mnsc.2025.00535

[15] Neil Dallaway, Samuel J. E. Lucas, and Christopher Ring. 2022. Cognitive tasks elicit mental fatigue and impair subsequent physical task endurance: Effects of task duration and type. *Psychophysiology* 59, 12 (December 2022). https://doi.org/10.1111/psyp.14126

[16] Empatica Inc. 2025. EmbracePlus. Retrieved July 29, 2025 from https://www.empatica.com/en-eu/embraceplus/

[17] Sarah Fakhoury, Aaditya Naik, Georgios Sakkas, Saikat Chakraborty, and Shuvendu K. Lahiri. 2024. LLM-Based Test-Driven Interactive Code Generation: User Study and Empirical Evaluation. *IEEE Transactions on Software Engineering* 50, 9 (September 2024), 2254–2268. https://doi.org/10.1109/TSE.2024.3428972

[18] Thomas Fritz, Andrew Begel, Sebastian C. Müller, Serap Yigit-Elliott, and Manuela Züger. 2014. Using psycho-physiological measures to assess task difficulty in software development. In *Proceedings - International Conference on Software Engineering*, May 31, 2014. IEEE Computer Society, 402–413. https://doi.org/10.1145/2568225.2568266

[19] Martin Gjoreski, Tine Kolenik, Timotej Knez, Mitja Luštrek, Matjaž Gams, Hristijan Gjoreski, and Veljko Pejović. 2020. Datasets for Cognitive Load Inference Using Wearable Sensors and Psychological

Traits. *Applied Sciences* 10, 11 (May 2020), 3843. https://doi.org/10.3390/app10113843

[20] Sandra G. Hart. 2006. Nasa-Task Load Index (NASA-TLX); 20 Years Later. *Proceedings of the Human Factors and Ergonomics Society Annual Meeting* 50, 9 (October 2006), 904–908. https://doi.org/10.1177/154193120605000909

[21] Hye-Geum Kim, Eun-Jin Cheon, Dai-Seg Bai, Young Hwan Lee, and Bon-Hoon Koo. 2018. Stress and Heart Rate Variability: A Meta-Analysis and Review of the Literature. *Psychiatry Investig.* 15, 3 (March 2018), 235–245. https://doi.org/10.30773/pi.2017.08.17

[22] Andre N. Meyer, Earl T. Barr, Christian Bird, and Thomas Zimmermann. 2021. Today Was a Good Day: The Daily Life of Software Developers. *IEEE Transactions on Software Engineering* 47, 5 (May 2021), 863–880. https://doi.org/10.1109/TSE.2019.2904957

[23] Andreas B. Neubauer, Stacey B. Scott, Martin J. Sliwinski, and Joshua M. Smyth. 2020. How was your day? Convergence of aggregated momentary and retrospective end-of-day affect ratings across the adult life span. *J. Pers. Soc. Psychol.* 119, 1 (July 2020), 185–203. https://doi.org/10.1037/pspp0000248

[24] Abi Noda, Margaret Anne Storey, Nicole Forsgren, and Michaela Greiler. 2023. DevEx: What Actually Drives Productivity. *Queue* 21, 2 (April 2023), 35–53. https://doi.org/10.1145/3595878

[25] Adrian M. Owen, Kathryn M. McMillan, Angela R. Laird, and Ed Bullmore. 2005. N-back working memory paradigm: A meta-analysis of normative functional neuroimaging studies. *Hum. Brain Mapp.* 25, 1 (May 2005), 46–59. https://doi.org/10.1002/hbm.20131

[26] Elise Paradis, Kate Grey, Quinn Madison, Daye Nam, Andrew Macvean, Vahid Meimand, Nan Zhang, Ben Ferrari-Church, and Satish Chandra. 2025. How Much Does AI Impact Development Speed? an Enterprise-Based Randomized Controlled Trial. In *2025 IEEE/ACM 47th International Conference on Software Engineering: Software Engineering in Practice (ICSE-SEIP)*, April 27, 2025. IEEE, 618–629. https://doi.org/10.1109/ICSE-SEIP66354.2025.00060

[27] Cornelia Setz, Bert Arnrich, Johannes Schumm, Roberto La Marca, Gerhard Tröster, and Ulrike Ehlert. 2010. Discriminating stress from cognitive load using a wearable eda device. *IEEE Transactions on Information Technology in Biomedicine* 14, 2 (March 2010), 410–417. https://doi.org/10.1109/TITB.2009.2036164

[28] Margaret-Anne Storey. 2026. From Technical Debt to Intent Debt: Rethinking Software Health in the Age of AI. *ArXiv* (2026). Retrieved August 10, 2026 from https://arxiv.org/abs/2603.22106

[29] J. R. Stroop. 1935. Studies of interference in serial verbal reactions. *J. Exp. Psychol.* 18, 6 (December 1935), 643–662. https://doi.org/10.1037/h0054651

[30] Ningzhi Tang, Meng Chen, Zheng Ning, Aakash Bansal, Yu Huang, Collin McMillan, and Toby Jia-Jun Li. 2024. Developer Behaviors in Validating and Repairing LLM-Generated Code Using IDE and Eye Tracking. In *2024 IEEE Symposium on Visual Languages and Human-Centric Computing (VL/HCC)*, September 02, 2024. IEEE, 40–46. https://doi.org/10.1109/VL/HCC60511.2024.00015

[31] Annemarie Uhlig, Charlotte Brandebusemeyer, Frédéric Li, Falco Lentzsch, Marcin Grzegorzek, and Bert Arnrich. 2026. Analyzing Cognitive Load of Software Developers Using Wearables During Their Everyday Work. *Manuscript submitted for review*.

[32] Priyan Vaithilingam, Tianyi Zhang, and Elena L. Glassman. 2022. Expectation vs. Experience: Evaluating the Usability of Code Generation Tools Powered by Large Language Models. In *CHI Conference on Human Factors in Computing Systems Extended Abstracts (CHI '22 Extended Abstracts)*, April 27, 2022. ACM, New York, NY, USA, 1–7. https://doi.org/10.1145/3491101.3519665

[33] Annie Vella and Kelly Blincoe. 2026. The Impact of AI Coding Assistants on Software Engineering: A Longitudinal Study. (May 2026). https://doi.org/10.48550/arXiv.2605.23135

[34] Christiaan H. Vinkers, Renske Penning, Juliane Hellhammer, Joris C. Verster, John H. G. M. Klaessens, Berend Olivier, and Cor J. Kalkman. 2013. The effect of stress on core and peripheral body temperature in humans. *Stress* 16, 5 (September 2013), 520–530. https://doi.org/10.3109/10253890.2013.807243

[35] Barbara Weber, Thomas Fischer, and René Riedl. 2021. Brain and autonomic nervous system activity measurement in software engineering: A systematic literature review. *Journal of Systems and Software* 178, (August 2021). https://doi.org/10.1016/j.jss.2021.110946

[36] Justin D. Weisz, Shraddha Vijay Kumar, Michael Muller, Karen-Ellen Browne, Arielle Goldberg, Katrin Ellice Heintze, and Shagun Bajpai. 2025. Examining the Use and Impact of an AI Code Assistant on Developer Productivity and Experience in the Enterprise. In *Proceedings of the Extended Abstracts of the CHI Conference on Human Factors in Computing Systems*, April 26, 2025. ACM, New York, NY, USA, 1–13. https://doi.org/10.1145/3706599.3706670